\documentclass[10pt,prd,twocolumn,fleqn,superscriptaddress,notitlepage,nofootinbib,preprintnumbers]{revtex4-1}
\pdfoutput=1
\usepackage{amsmath,amssymb,graphicx,xspace,listings,booktabs,comment,multirow,nicefrac,xcolor,nicefrac,array}
\usepackage[squaren]{SIunits}
\usepackage[hidelinks]{hyperref}
\newcommand{\BlackHat}{B\protect\scalebox{0.8}{LACK}H\protect\scalebox{0.8}{AT}\xspace}
\newcommand{\MCFM}{M\protect\scalebox{0.8}{CFM}\xspace}
\newcommand{\MadLoop}{M\protect\scalebox{0.8}{AD}L\protect\scalebox{0.8}{OOP}}
\newcommand{\OpenLoops}{O\protect\scalebox{0.8}{PEN}L\protect\scalebox{0.8}{OOPS}}
\newcommand{\Pythia}{P\protect\scalebox{0.8}{YTHIA}\xspace}
\newcommand{\Rambo}{R\protect\scalebox{0.8}{AMBO}\xspace}
\newcommand{\Recola}{R\protect\scalebox{0.8}{ECOLA}}
\newcommand{\Sherpa}{S\protect\scalebox{0.8}{HERPA}\xspace}
\newcommand{\Vincia}{V\protect\scalebox{0.8}{INCIA}\xspace}
\newcommand{\Order}[1]{\ensuremath{\mathcal{O}\left(#1\right)}}

\renewcommand{\minute}{\mathrm{m}}
\newcolumntype{C}[1]{>{\centering\arraybackslash}p{#1}}
\AtBeginDocument{
  \heavyrulewidth=.08em
  \lightrulewidth=.05em
  \cmidrulewidth=.03em
  \belowrulesep=.65ex
  \belowbottomsep=0pt
  \aboverulesep=.4ex
  \abovetopsep=0pt
  \cmidrulesep=\doublerulesep
  \cmidrulekern=.5em
  \defaultaddspace=.5em
}
\definecolor{light-gray}{gray}{0.95}
\begin{document}
\title{Accelerating LHC phenomenology with analytic one-loop amplitudes:\\A C++ interface to \MCFM}
\preprint{FERMILAB-PUB-21-290-T, MCNET-21-13}
\author{John M Campbell}
\affiliation{Fermi National Accelerator Laboratory, Batavia, IL, 60510, USA}
\author{Stefan H{\"o}che}
\affiliation{Fermi National Accelerator Laboratory, Batavia, IL, 60510, USA}
\author{Christian T Preuss}
\affiliation{School of Physics and Astronomy, Monash University, Wellington Road, Clayton, VIC-3800, Australia}
\begin{abstract}
  The evaluation of one-loop matrix elements is one of the main bottlenecks in precision calculations
  for the high-luminosity phase of the Large Hadron Collider. To alleviate this problem, a new C++ interface
  to the \MCFM parton-level Monte Carlo is introduced, giving access to an extensive library of analytic results
  for one-loop amplitudes. Timing comparisons are presented for a large set of Standard Model processes.
  These are relevant for high-statistics event simulation in the context of experimental analyses
  and precision fixed-order computations.
\end{abstract}
\maketitle

\section{Introduction}
Many measurements at particle colliders can only be made with the help of precise Standard Model predictions,
which are typically derived using fixed-order perturbation theory at the next-to-leading order (NLO)
or next-to-next-to-leading order (NNLO) in the strong and/or electroweak coupling. Unitarity-based techniques
and improvements in tensor reduction during the past two decades have enabled the computation of many
new one-loop matrix elements, often using fully numeric techniques~\cite{Berger:2008sj,Berger:2009ep,Berger:2010vm,Berger:2010zx,Ita:2011wn,Bern:2013gka,Hirschi:2011pa,Alwall:2014hca,Cascioli:2011va,Buccioni:2017yxi,Buccioni:2019sur,Badger:2010nx,Badger:2012pg,Cullen:2011ac,Cullen:2014yla,Actis:2012qn,Actis:2016mpe,Denner:2017vms,Denner:2017wsf}.
The algorithmic appeal and comparable simplicity of the novel approaches has also led to the partial automation
of the computation of one-loop matrix elements in arbitrary theories, including effective field theories
that encapsulate the phenomenology of a broad range of additions to the
Standard Model~\cite{Degrande:2014vpa,Degrande:2020evl}.
With this ``NLO revolution'' precision phenomenology has entered a new era.

It has become clear, however, that the fully numeric computation of one-loop matrix elements is
not without its drawbacks, the most relevant being a relatively large computational complexity.
While the best methods exhibit good scaling with the number of final-state particles and are
the only means to perform very high multiplicity calculations, it is prudent to resort to known
analytic results whenever they are available and computational resources are scarce. 
The problem has become pressing due to the fact that the computing power on the
Worldwide LHC Computing Grid (WLCG) is projected to fall short of the demand by at least a factor two
in the high-luminosity phase of the Large Hadron Collider (LHC)~\cite{Alves:2017she,Buckley:2019kjt,Valassi:2020ueh,Aarrestad:2020ngo}. Moreover, most techniques for fully differential
NNLO calculations rely on the fast and numerically stable evaluation of one-loop results in
infrared-singular regions of phase space, further increasing the demand for efficient
one-loop computations~\cite{Boughezal:2016wmq,Campbell:2019dru}.

In this letter, we report on an extension of the well-known NLO parton-level program
\MCFM~\cite{Campbell:1999ah,Campbell:2011bn,Campbell:2015qma,Campbell:2019dru}, which allows
the one-loop matrix elements in \MCFM to be accessed using the Binoth Les Houches Accord
(BLHA)~\cite{Binoth:2010xt,Alioli:2013nda} via a direct C++ interface\footnote{
  The source code is available at \href{https://gitlab.com/mcfm-team/releases}{gitlab.com/mcfm-team/releases}.}\edef\tmpfn{\the\value{footnote}}.
This is in the same spirit as the BLHA interface to the \BlackHat library~\cite{Berger:2008sj},
which gives access to analytic matrix elements for $V+$jet(s), $\gamma\gamma$(+jet) and di-(tri-)jet production.
We have constructed the new interface for the most relevant Standard-Model processes
available in \MCFM, representing a selection of $2 \to n$ processes with $n \leq 4$.
As a proof of generality, we have implemented it in the \Sherpa \cite{Sherpa:2019gpd}
and \Pythia \cite{Sjostrand:2014zea} event generation frameworks\footnote{The \Pythia version has been tested in the context of NLO matrix-element corrections (cf.~\cite{Hartgring:2013jma,Baberuxki:2019ifp}) in the \Vincia shower \cite{Brooks:2020upa}. The implementation of NLO MECs in \Vincia and the \MCFM interface are planned to be made public in a future \Pythia~8.3 release.}\edef\vncfn{\the\value{footnote}}.
We test the newly developed methods in both a stand-alone setup and a typical setup of the \Sherpa event generator,
and summarize the speed gains in comparison to automated one-loop programs.

\section{Available processes}
The Standard Model processes currently available through the \MCFM one-loop interface
are listed in Tab.~\ref{tab:SM_Procs}, with additional processes available in the Higgs effective theory
shown in Tab.~\ref{tab:HEFT_Procs}. All processes are implemented in a crossing-invariant fashion.
As well as processes available in the most recent version
of the \MCFM code (v10.0), the interface also allows access to previously unreleased
matrix elements for $pp\to\gamma j j$~\cite{Campbell:2016lzl} and di-jet production.
Further processes listed in the \MCFM manual~\cite{mcfm} may be included upon request.

In assembling the interface we have modified the original \MCFM routines such that, as far as possible, overhead
associated with the calculation of all partonic channels -- as required for the normal operation of the \MCFM code --
is avoided, and only the specific channel that is requested is computed.  Additionally, all matrix elements are
calculated using the complex-mass scheme~\cite{Denner:1999gp,Denner:2005fg} and a non-diagonal
form of the CKM matrix may be specified in the interface.  In general, effects due to loops containing
a massive top quark are fully taken into account, with the additional requirement that the width of the
top quark is set to zero.\footnote{An approximate form for top-quark loops is used for the
processes $q q^\prime \to \gamma q q^\prime$, $q g \to \gamma q g$, $q \bar q \to e^- e^+ h g$ and
$gg \to gg$, so that strict agreement with other OLPs for these processes requires the removal of the top-quark loops
in those.}
  The intent is that the interface can therefore be used as a direct replacement for a numerical
one-loop provider (OLP).  We have checked, on a point-by-point basis, that the one-loop matrix elements returned
by the interface agree perfectly with those provided by \OpenLoops2, \Recola2 and \MadLoop5.
A brief overview of the structure of the interface is given in Appendix~\ref{sec:app}.

\begin{table}[t]
	\centering
	\caption{Processes available in the Standard Model.}
	\begin{tabular}{p{0.15\textwidth}C{0.07\textwidth}C{0.07\textwidth}p{0.07\textwidth}}\toprule
		Process & Order EW & Order QCD & Reference\\ \midrule
		$p p \to \ell^+ \ell^-$ & 2 & 1 & -- \\
		$p p \to \ell^+ \ell^- j$ & 2 & 2 & \cite{Bern:1997sc,Campbell:2016tcu} \\
		$p p \to \ell^+ \ell^- j j$ & 2 & 3 & \cite{Bern:1997sc,Campbell:2016tcu} \\ \midrule
		$p p \to \ell^\pm \nu_\ell$ & 2 & 1 & -- \\
		$p p \to \ell^\pm \nu_\ell j$ & 2 & 2 & \cite{Bern:1997sc,Campbell:2016tcu} \\
		$p p \to \ell^\pm \nu_\ell j j$ & 2 & 3 & \cite{Bern:1997sc,Campbell:2016tcu} \\ \midrule
		$p p \to h_0$ & 1 & 2 & -- \\
		$p p \to h_0 j$ & 1 & 3 & \cite{Ellis:1987xu} \\
		$p p \to h_0 j j$ & 1 & 4 & \cite{Ellis:2018hst,Budge:2020oyl} \\ \midrule
		$p p \to h_0 h_0$ & 2 & 2 & \cite{Glover:1987nx} \\ \midrule
		$p p \to \ell^+ \ell^- h_0$ & 3 & 1 & -- \\
		$p p \to \ell^+ \ell^- h_0 j$ & 3 & 2 & \cite{Campbell:2016jau} \\ \midrule
		$p p \to \ell^\pm \nu_\ell h_0$ & 3 & 1 & -- \\
		$p p \to \ell^\pm \nu_\ell h_0 j$ & 3 & 2 & \cite{Campbell:2016jau} \\ \midrule
		$p p \to \gamma j$ & 1 & 2 & \cite{Ellis:1980wv,Aurenche:1986ff} \\
		$p p \to \gamma jj$ & 1 & 3 & \cite{Campbell:2016lzl} \\ \midrule
		$p p \to \gamma \gamma$ & 2 & 1 & -- \\
		$g g \to \gamma \gamma$ & 2 & 2 & \cite{Campbell:2016yrh} \\
		$p p \to \gamma \gamma j$ & 2 & 2 & \cite{Campbell:2014yka} \\ \midrule
		$p p \to \gamma \gamma \gamma$ & 3 & 1 & \cite{Campbell:2014yka} \\\midrule
		$p p \to \gamma \gamma \gamma \gamma$ & 4 & 1 & \cite{Dennen:2014vta} \\\midrule
		$p p \to \ell^+ \ell^- \gamma$ & 3 & 1 & \cite{Dixon:1998py,Campbell:1999ah} \\
		$p p \to \ell^\pm \nu_\ell \gamma$ & 3 & 1 & \cite{Dixon:1998py,Campbell:1999ah} \\
		$p p \to \nu_\ell \bar{\nu}_\ell \gamma$ & 3 & 1 & \cite{Dixon:1998py,Campbell:1999ah} \\ \midrule
		$p p \to \ell^+ \ell^{\prime -} \nu_\ell \bar{\nu}_{\ell'}$ & 4 & 1 & \cite{Dixon:1998py,Campbell:1999ah} \\
		$p p \to \ell^+ \ell^- \nu_{\ell'} \bar{\nu}_{\ell'}$ & 4 & 1 & \cite{Dixon:1998py,Campbell:1999ah} \\
		$p p \to \ell^+ \ell^- \ell^{\prime\,+} \ell^{\prime\,-}$ & 4 & 1 & \cite{Dixon:1998py,Campbell:1999ah} \\
		$p p \to \ell^+ \ell^+ \ell^- \ell^-$ & 4 & 1 & \cite{Dixon:1998py,Campbell:1999ah} \\ \midrule
		$p p \to \ell^+ \ell^-\ell^{\prime\pm} \nu_{\ell'}$ & 4 & 1 & \cite{Dixon:1998py,Campbell:1999ah} \\
		$p p \to \ell^\pm \nu_\ell \nu_{\ell'} \bar{\nu}_{\ell'}$ & 4 & 1 & \cite{Dixon:1998py,Campbell:1999ah} \\ \midrule
		$p p \to t \bar{t}$ & 0 & 3 & \cite{Nason:1989zy} \\ \midrule
		$p p \to j j$ & 0 & 3 & \cite{Ellis:1985er}
		\\ \bottomrule
	\end{tabular}
	\label{tab:SM_Procs}
\end{table}
\begin{table}[t]
	\centering
	\caption{Processes available in the Higgs EFT.}
	\begin{tabular}{p{0.15\textwidth}C{0.07\textwidth}C{0.07\textwidth}p{0.07\textwidth}}\toprule
		Process & Order EW & Order QCD & Reference \\ \midrule
		$p p \to h_0$ & 1 & 2 & -- \\
		$p p \to h_0 j$ & 1 & 3 & \cite{Schmidt:1997wr} \\
		$p p \to h_0 j j$ & 1 & 4 & \cite{Dixon:2004za,Ellis:2005qe,Badger:2006us,Badger:2007si,Glover:2008ffa,Badger:2009hw,Dixon:2009uk,Badger:2009vh} \\ \bottomrule
	\end{tabular}
	\label{tab:HEFT_Procs}
\end{table}

\section{Timing benchmarks}
\begin{figure}[t]
  \centering
  \includegraphics[width=0.4875\textwidth]{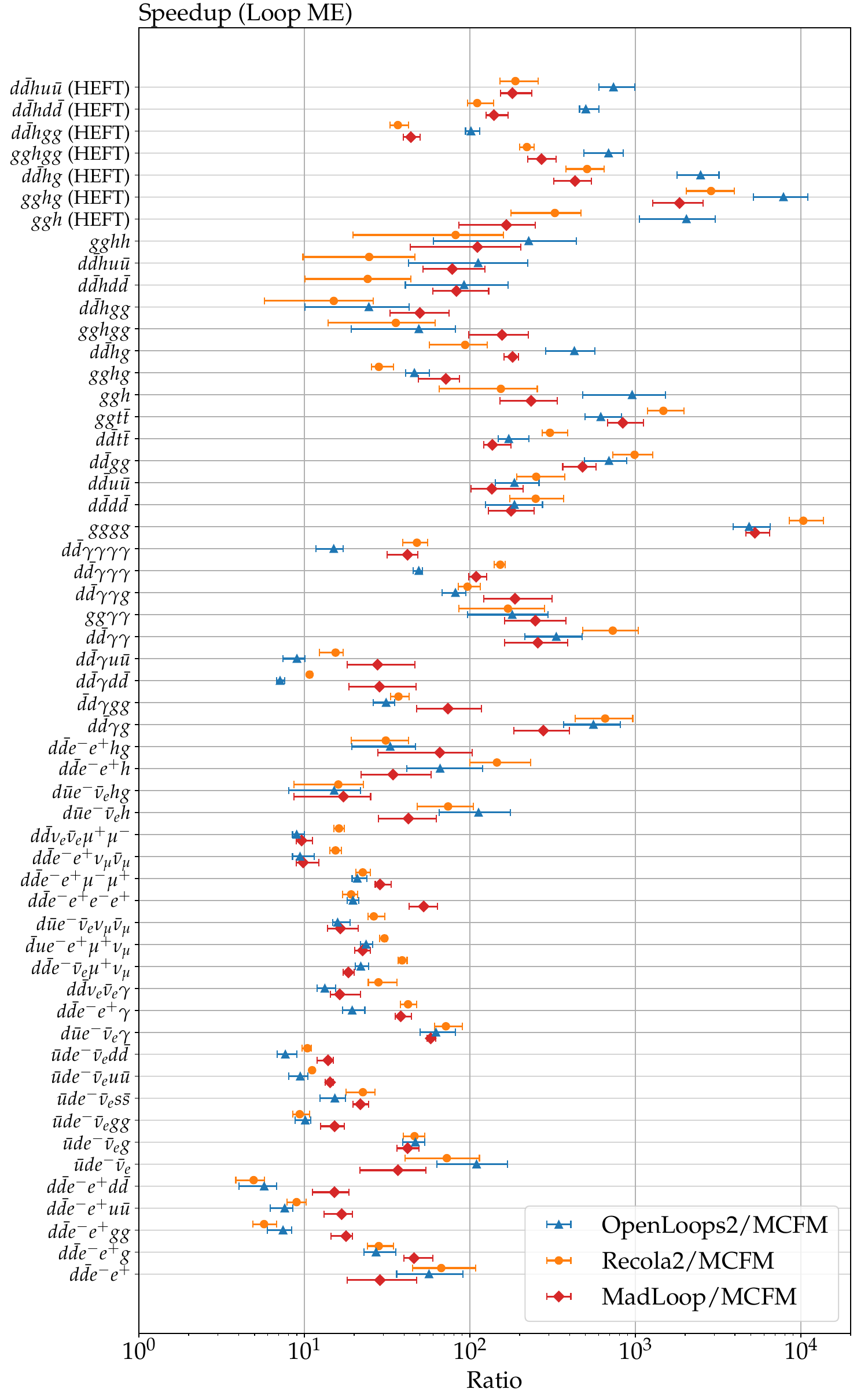}
  \caption{CPU time ratio of \OpenLoops, \Recola, and \MadLoop5 to \MCFM at the level of loop matrix elements.}
  \label{fig:runtimeRatioME}
\end{figure}

To gauge the efficiency gains compared to automated one-loop providers, we
compare the evaluation time in \MCFM against \OpenLoops2, \Recola2, and \MadLoop5 using their
default settings. In particular, we neither tune nor deactivate their stability systems. 
The tests are conducted in three stages. First, we test the CPU time needed for the evaluation
of loop matrix elements at single phase space points; in a second stage, we test the speedup
in the calculation of Born-plus-virtual contributions of NLO calculations using realistic setups;
lastly, we compare the CPU time of the different OLPs in a realistic multi-jet merged calculation.
In all cases, we estimate the dependence on the computing hardware by running all tests on a total
of four different CPUs, namely
\begin{itemize}
\item Intel$^\text{\textregistered}$ Xeon$^\text{\textregistered}$ E5-2650 v2 (2.60GHz, 20MB)
\item Intel$^\text{\textregistered}$ Xeon$^\text{\textregistered}$ Gold 6150 (2.70GHz, 24.75MB)
\item Intel$^\text{\textregistered}$ Xeon$^\text{\textregistered}$ Platinum 8260 (2.40GHz, 35.75MB)
\item Intel$^\text{\textregistered}$ Xeon Phi$^\text{\texttrademark}$ 7210 (1.30GHz, 32MB)
\end{itemize}
For the timing tests at matrix-element level, we use stand-alone interfaces to the respective tools
and sample phase space points flatly using the \Rambo algorithm \cite{Kleiss:1985gy}.
We do not include the time needed for phase-space point generation in our results, and we
evaluate a factor 10 more phase-space points in \MCFM in order to obtain more accurate timing
measurements at low final-state multiplicity.
The main programs and scripts we used for this set of tests are publicly available\footnotemark[\tmpfn].
The results are collected in Fig.~\ref{fig:runtimeRatioME}, where we show all distinct partonic
configurations that contribute to the processes listed in Tabs.~\ref{tab:SM_Procs} and~\ref{tab:HEFT_Procs}.
We use the average across the different CPUs as the central value, while the error bars range from the minimal to the maximal value. 
The interface to \MCFM typically evaluates matrix elements a factor $10$--$100$ faster than the numerical
one-loop providers, although for a handful of (low multiplicity) cases this factor can be in the 1,000--10,000
range.

\begin{figure}[t]
  \centering
  \includegraphics[width=0.4875\textwidth]{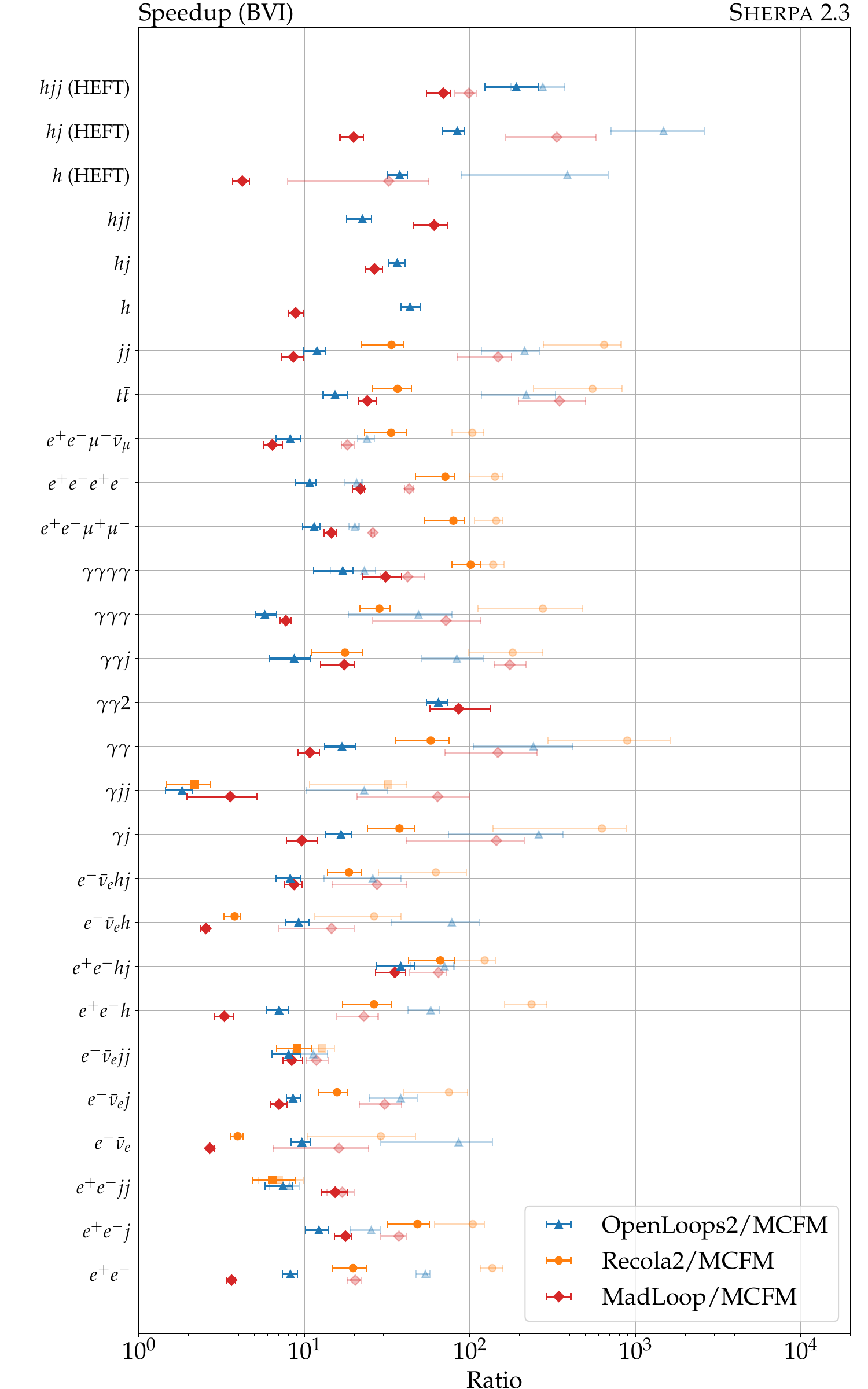}
  \caption[]{CPU time ratio of \OpenLoops2, \Recola2~$^\text{\scriptsize{5}}$, and \MadLoop5 to \MCFM at the level of
    Born-like contributions to the NLO cross section (BVI).}
  \label{fig:runtimeRatioBVI}
\end{figure}

We perform a second set of tests, using the \Sherpa event generator~\cite{Gleisberg:2008ta,Sherpa:2019gpd},
its existing OLP interfaces to \OpenLoops2 and \Recola2~\footnote{\label{fn:rec}At the time of this study, \Sherpa provided an interface to \Recola's Standard Model implementation only.}~\footnote{For $V+2j$ processes, we use \Recola1 due to compatibility issues with \Recola2.}\edef\recfn{\the\value{footnote}} \cite{Biedermann:2017yoi},
and a dedicated interface to \MadLoop5~\footnote{We thank Valentin Hirschi for his help in constructing a dedicated \MadLoop5 interface to \Sherpa. This interface will be described in detail elsewhere.}.
With these interfaces we test the speedup in the calculation of the Born-like contributions to a typical
NLO computation for the LHC at $\sqrt{s}=14$~TeV, involving the loop matrix elements
in Tabs.~\ref{tab:SM_Procs} and~\ref{tab:HEFT_Procs}.
The scale choices and phase-space cuts used in these calculations are listed in App.~\ref{app:cuts_for_bvi}.
Figure~\ref{fig:runtimeRatioBVI} shows the respective timing ratios. It is apparent that the large gains
observed in Fig.~\ref{fig:runtimeRatioME} persist in this setup, because the Born-like contributions to
the NLO cross section consist of the Born, integrated subtraction terms, collinear mass factorization
counterterms and virtual corrections (BVI), and the timing is dominated by the loop matrix elements
if at least one parton is present in the final state at Born level.
The usage of \MCFM speeds up the calculation by a large factor compared to the automated OLPs,
with the exception of very simple processes, such as $pp\to\ell\bar{\ell}$, $pp\to h$, etc., where the overhead
from process management and integration in Sherpa dominates. To assess this overhead we also compute
the timing ratios after subtracting the time that the Sherpa computation would take without a loop matrix element.
The corresponding results are shown in a lighter shade and confirm that the Sherpa overhead is significant
at low multiplicity and becomes irrelevant at higher multiplicity.

In the final set of tests we investigate a typical use case in the context of parton-level
event generation for LHC experiments. We use the \Sherpa event generator in a multi-jet merging setup
for $pp\to W$+jets and $pp\to Z$+jets~\cite{Hoeche:2012yf} at $\sqrt{s}=8$~TeV,
with a jet separation cut of $Q_{\rm cut}=20$~GeV,
and a maximum number of five final state jets at the matrix-element level. Up to two-jet final states
are computed at NLO accuracy. In this use case, the gains observed
in Figs.~\ref{fig:runtimeRatioME} and~\ref{fig:runtimeRatioBVI} will be greatly diminished, because
the timing is dominated by the event generation efficiency for the highest multiplicity tree-level
matrix elements~\cite{Hoeche:2019rti} and influenced by particle-level event generation as
well as the clustering algorithm needed for multi-jet merging%
\footnote{In this study we do not address the question of additional timing overhead due to NLO electroweak corrections or PDF reweighting \cite{Bothmann:2016nao}, which could both be relevant in practice. It has recently been shown that in good implementations of the reweighting and EW correction algorithm, the additional overhead will not be sizable~\cite{hsfdocument}.}.
We make use of the efficiency improvements described in Ref.~\cite{Danziger:2020aa}, in particular
neglecting color and spin correlations in the S-MC@NLO matching procedure~\cite{Hoeche:2011fd}.
We do not include underlying event simulation or hadronization.
The results in Tab.~\ref{tab:runtimeRatioMerged}
still show a fairly substantial speedup when using \MCFM. We point out that a higher gain could be achieved
by also making use of \MCFM's implementation of analytic matrix elements for real-emission corrections
and Catani-Seymour dipole terms.

\begin{table}[t]
\centering
\caption{CPU time ratios in an NLO multi-jet merged setup using \Sherpa.}
\label{tab:runtimeRatioMerged}
\begin{tabular}{lcc}\toprule
  Merged Process & \Sherpa$+$ & \Sherpa$+$ \\ 
  {\scriptsize$n \leq 2~@~\mathrm{NLO}$} & \multirow{2}{*}{\footnotesize\nicefrac{\OpenLoops2}{\MCFM}} & \multirow{2}{*}{\footnotesize\nicefrac{\MadLoop5}{\MCFM}} \\
  {\scriptsize$n \leq 5~@~\mathrm{LO}$} & & \\ \midrule
  $pp\to Z+nj$ & $1.83^{+0.20}_{-0.12}$ & $3.01^{+0.26}_{-0.18}$ \\ \midrule
  $pp \to W^++nj$ & $1.34^{+0.06}_{-0.07}$ & $1.36^{+0.03}_{-0.03}$ \\ \midrule
  $pp \to W^-+nj$ & $1.38^{+0.06}_{-0.04}$ & $1.38 ^{+0.07}_{-0.11}$ \\ \bottomrule
\end{tabular}
\end{table}

We close this section with a direct comparison of the CPU time needed for the calculation of
Drell-Yan processes with one and two jets using \Sherpa and \MCFM, up to a target precision on the
integration of $0.1$\% (one jet) or $0.3$\% (two jets). The center-of-mass energy
is $\sqrt{s}=14$~TeV, and the scale choices and cuts are listed in App.~\ref{app:cuts_for_bvi}.
The results are shown in Tab.~\ref{tab:intTimesSherpaMCFM}.
As might be expected when comparing a dedicated parton-level code with a general-purpose particle-level generator,
\MCFM is substantially faster than \Sherpa for the evaluation of all contributions to the NLO calculation.
These results indicate a few avenues for further improvements of general-purpose event generators.
With the efficient evaluation of virtual contributions in hand, attention should now turn to the calculation
of real-radiation configurations -- that represent the bottleneck for both \Sherpa and \MCFM.
In the simplest cases with up to 5 partons, the real radiation and dipole counterterms could be evaluated
using analytic rather than numerical matrix elements, by a suitable extension of the interface we have presented here.
In addition, the form of the phase-space generation may be improved for Born-like phase-space integrals.
Table~\ref{tab:intTimesSherpaMCFM} lists the number of phase-space points before cuts that are required
to achieve the target accuracy. We find that \MCFM uses fewer than half of the points needed by \Sherpa
in the Born-like phase-space integrals, while \Sherpa uses fewer points than \MCFM in the real-emission
type integrals but at a much higher computational cost. This confirms that \Sherpa's event generation is
indeed impaired by the slow evaluation of real-emission type matrix elements, and by the factorial scaling
of the diagram-based phase-space integration technique~\cite{Byckling:1969sx,Byckling:1970wn} used in its
calculations\footnote{We do not make use of \Sherpa's recursive phase-space generator~\cite{Gleisberg:2008fv},
  because it is available for color-sampled matrix element evaluation only. Color sampling would further
  reduce the efficiency of the integration, because the processes at hand involve a relatively small
  number of QCD partons.}.

\begin{table}[t]
\centering
\caption{Comparison of integration times using \Sherpa and \MCFM.}
\label{tab:intTimesSherpaMCFM}
\begin{tabular}{ll@{~}c@{~}c}\toprule
  Process & ~ & \Sherpa & \MCFM \\
  \scriptsize MC accuracy & & \scriptsize time / \#pts & \scriptsize time / \#pts \\ \midrule
  $pp \to Zj$ & {\scriptsize Born-like} & $76.8~\minute$ / 11.3M & $13.6~\minute$ /4.5M \\
  0.1\% & {\scriptsize real-like} & $38~\hour$ / 33.1M & $51.5~\minute$ / 22.5M \\ \midrule
  $pp \to Zjj$ & {\scriptsize Born-like} & $96.0~\hour$ / 22.4M & $19.6~\hour$ / 4.5M \\
  0.3\% & {\scriptsize real-like} & $830.4~\hour$ / 58.7M & $62.9~\hour$ / 83.8M \\ \midrule
  $pp \to W^+j$ & {\scriptsize Born-like} & $40.5~\minute$ / 12.8M & $7.37~\minute$ / 4.5M \\
  0.1\% & {\scriptsize real-like} & $16.9~\hour$ / 38.3M & $59.4~\minute$ / 36.0M \\ \midrule
  $pp \to W^+jj$ & {\scriptsize Born-like} & $14.1~\hour$ / 20.3M & $9.32~\hour$ / 7.2M \\
  0.3\% & {\scriptsize real-like} & $222.1~\hour$  / 38.9M & $54.4~\hour$ / 119.8M \\ \midrule
  $pp \to W^-j$ & {\scriptsize Born-like} & $34.1~\minute$ / 11.0M & $7.46~\minute$ / 4.5M \\
  0.1\% & {\scriptsize real-like} & $15.9~\hour$ / 40.5M & $47.2~\minute$ / 28.1M \\ \midrule
  $pp \to W^-jj$ & {\scriptsize Born-like} & $12.8~\hour$ / 20.0M & $7.34~\hour$ / 5.6M \\
  0.3\% & {\scriptsize real-like} & $281.1~\hour$ / 52.0M & $38.8~\hour$ / 83.8M \\ \bottomrule
\end{tabular}
\end{table}

\section{Numerical Stability}
As alluded to above, the numerical stability of one-loop amplitudes is of vital importance for both NLO and NNLO calculations, where the latter case necessitates a stable evaluation in single-unresolved phase-space regions. Here we wish to limit the discussion to this case and estimate the accuracy that can be expected from the one-loop amplitudes with an additional parton with respect to the Born multiplicity, i.e., those processses that correspond to the real-virtual contribution in an NNLO calculation.
To this end, we generate trajectories into the singular limits according to dipole kinematics, rescaling the Catani-Seymour variables of an initially hard configuration as
\begin{equation}
  y_{ijk} \to \begin{cases}
    \lambda \frac{s}{Q^2} & \mathrm{final-final} \\
    \frac{1}{1+\frac{Q^2(1-y_{ijk})}{\lambda s}} & \mathrm{final-initial} \\
    -z_i\frac{\lambda}{1-z_i\lambda} & \mathrm{initial-final} 
  \end{cases} \, ,
  \quad z_{i} \to z_{i}
  \label{eq:rescaleColl}
\end{equation}
in the collinear limit, and 
\begin{align}
  y_{ijk} & \to \begin{cases}
    \frac{C(1-z_{i})}{z_{i}} & \mathrm{final-final} \\
    \mathrm{sign}(y_{ijk})C\frac{1-z_{i}}{z_{i}} & \mathrm{initial-initial}
  \end{cases} \, , \nonumber\\
  \quad z_{i} & \to \begin{cases}
    1-2\frac{\lambda}{1+C} & \mathrm{final-final}  \\
    \frac{1}{1+2\frac{\lambda^2-\lambda\sqrt{1+\lambda^2}}{1+C}} & \mathrm{initial-initial}
  \end{cases} \, ,  \label{eq:rescaleSoft} \\
  C &= \frac{y_{ijk}z_{i}}{1-z_{i}} \nonumber
\end{align}
in the soft limit.
To assess the stability of the interface, we calculate the number of stable digits as
\begin{equation}
  N_\text{sd} = -\log_{10}\left(\frac{\vert V-V'\vert}{V}\right) \, ,
  \label{eq:numAcc}
\end{equation}
where $V$ and $V'$ denote the finite parts of the one-loop amplitude evaluated on two phase-space points that are rotated with respect to each other.

\begin{figure*}
  \centering
  \includegraphics[width=0.32\textwidth]{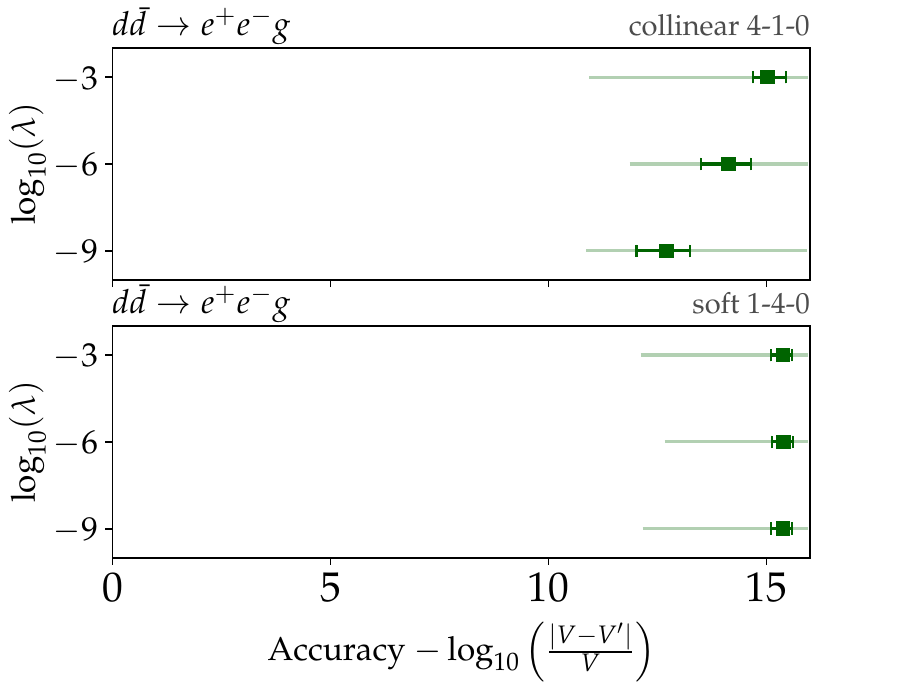}
  \includegraphics[width=0.32\textwidth]{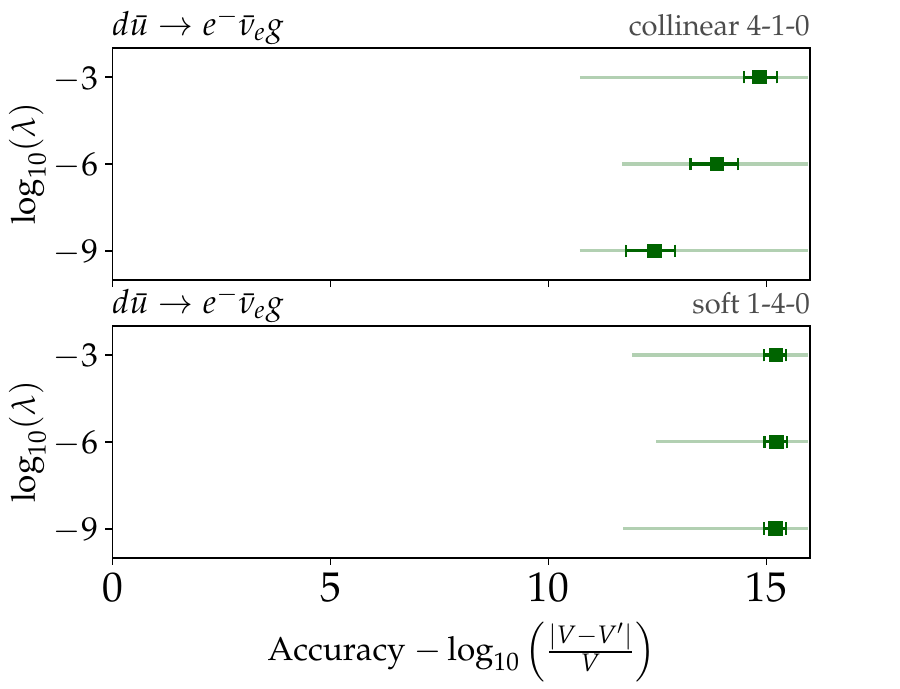}
  \includegraphics[width=0.32\textwidth]{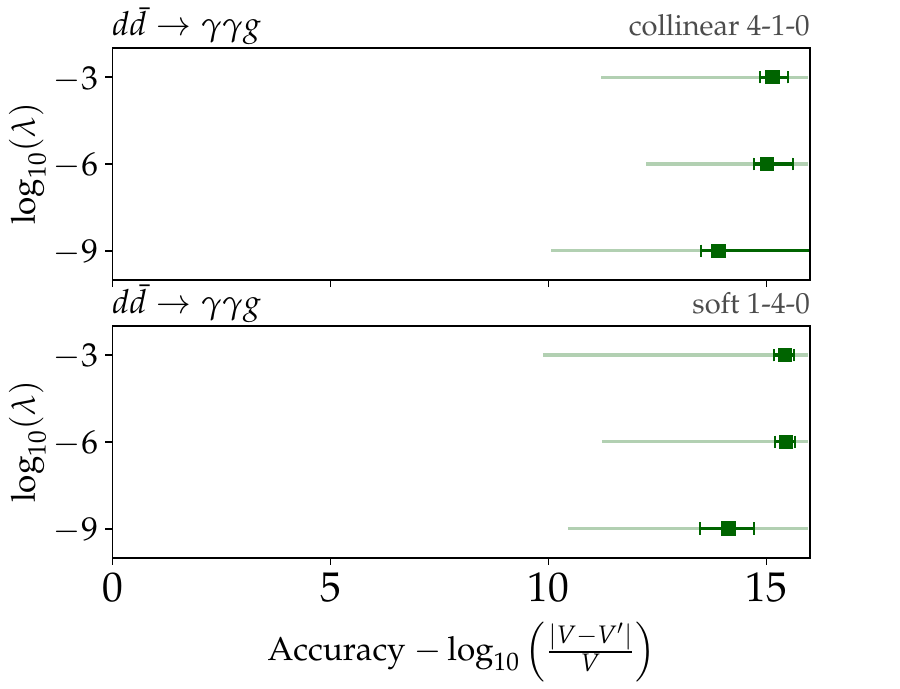}
  \caption[]{Test of the numerical accuracy of standard-model three-parton one-loop amplitudes in the soft and collinear limits.}
  \label{fig:scalingtestSM3j}
\end{figure*}

\begin{figure*}
  \centering
  \includegraphics[width=0.32\textwidth]{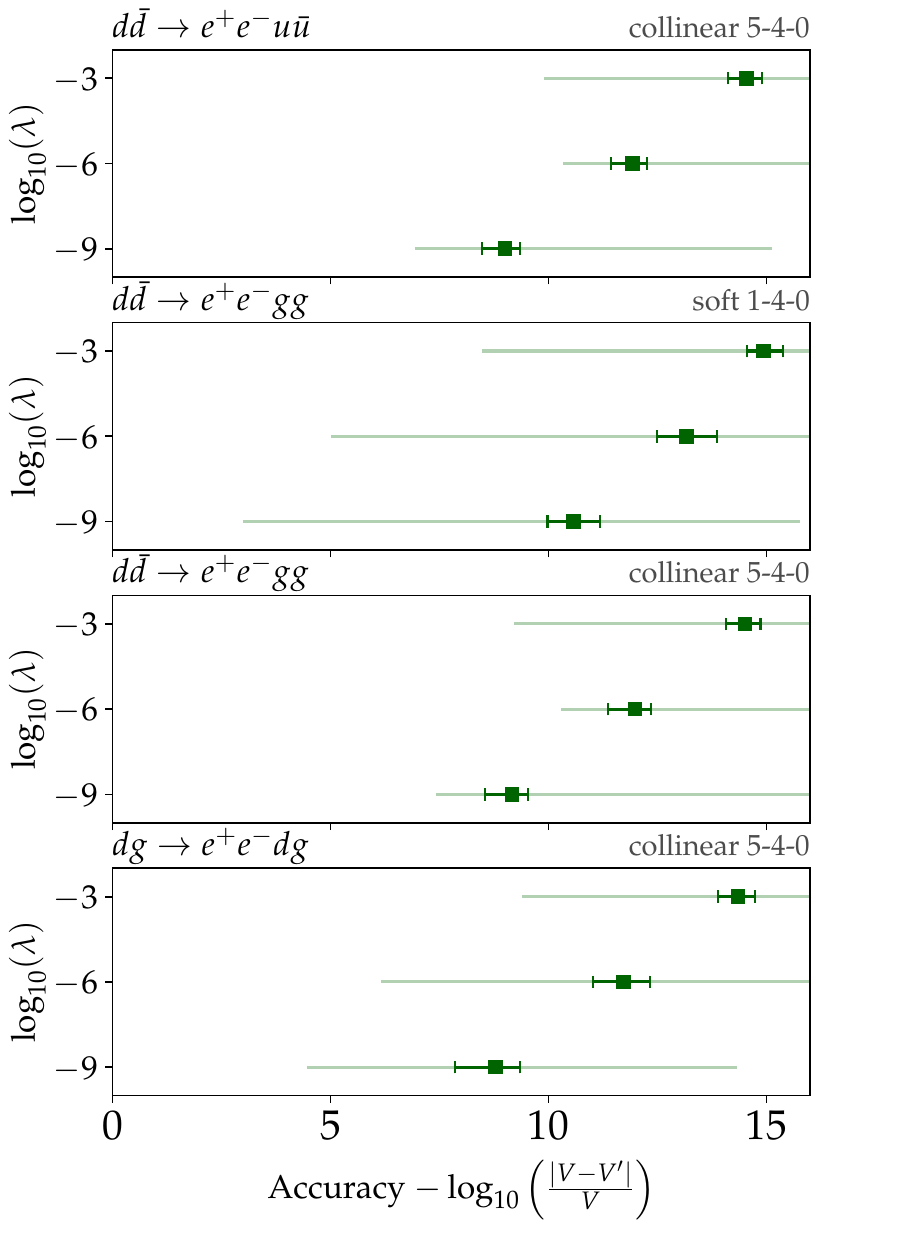}
  \includegraphics[width=0.32\textwidth]{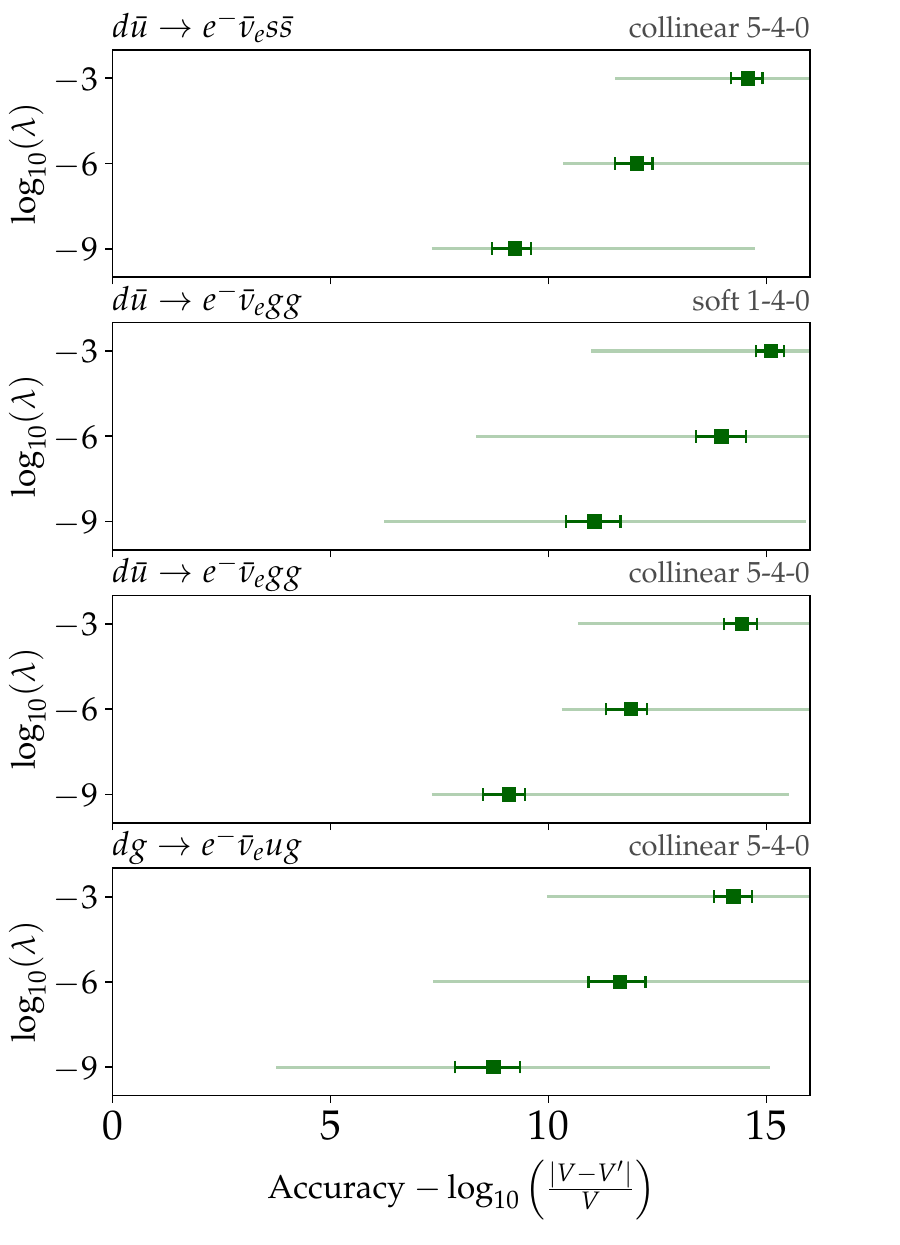}
  \includegraphics[width=0.32\textwidth]{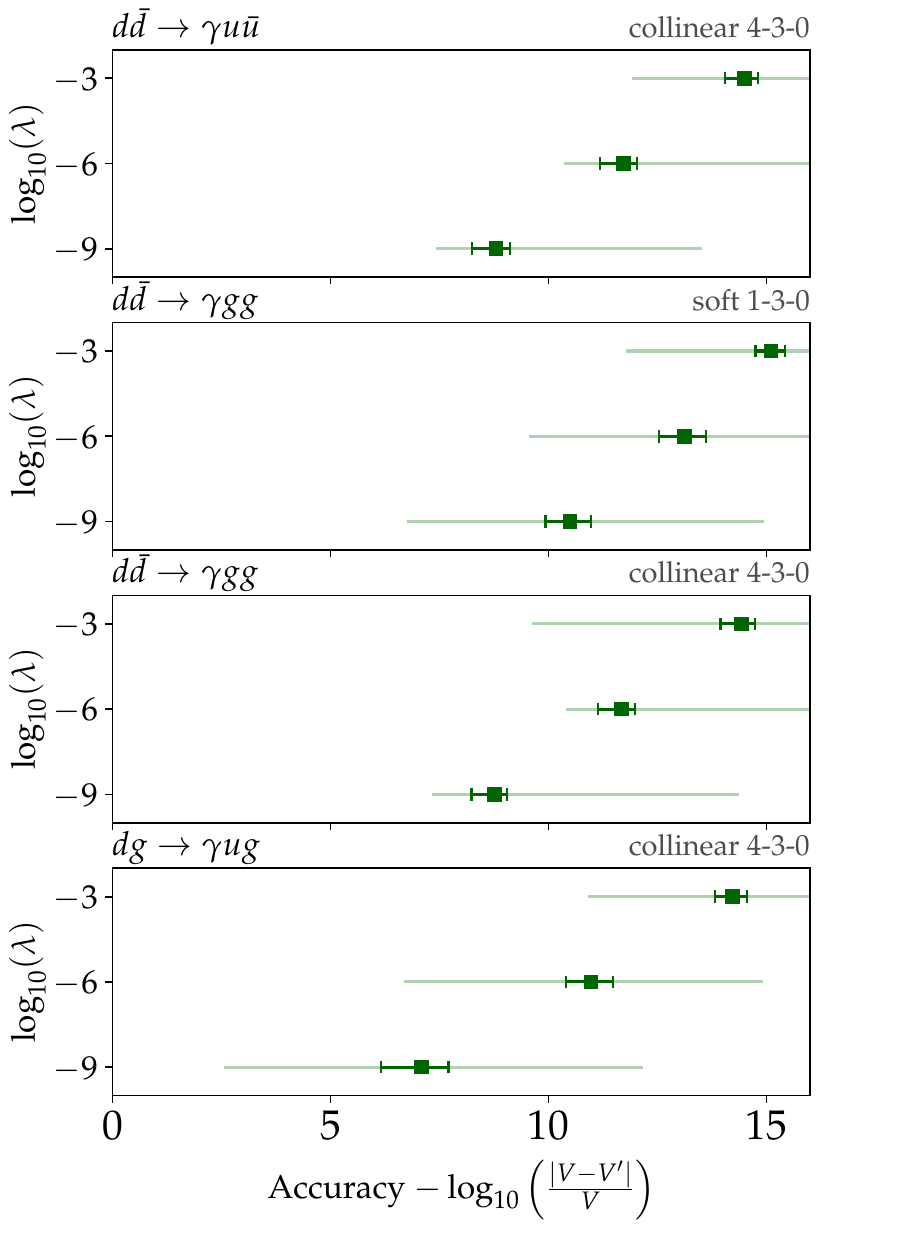}
  \caption[]{Test of the numerical accuracy of standard-model four-parton one-loop amplitudes in the soft and collinear limits.}
  \label{fig:scalingtestSM4j}
\end{figure*}

\begin{figure*}
  \centering
  \includegraphics[width=0.32\textwidth]{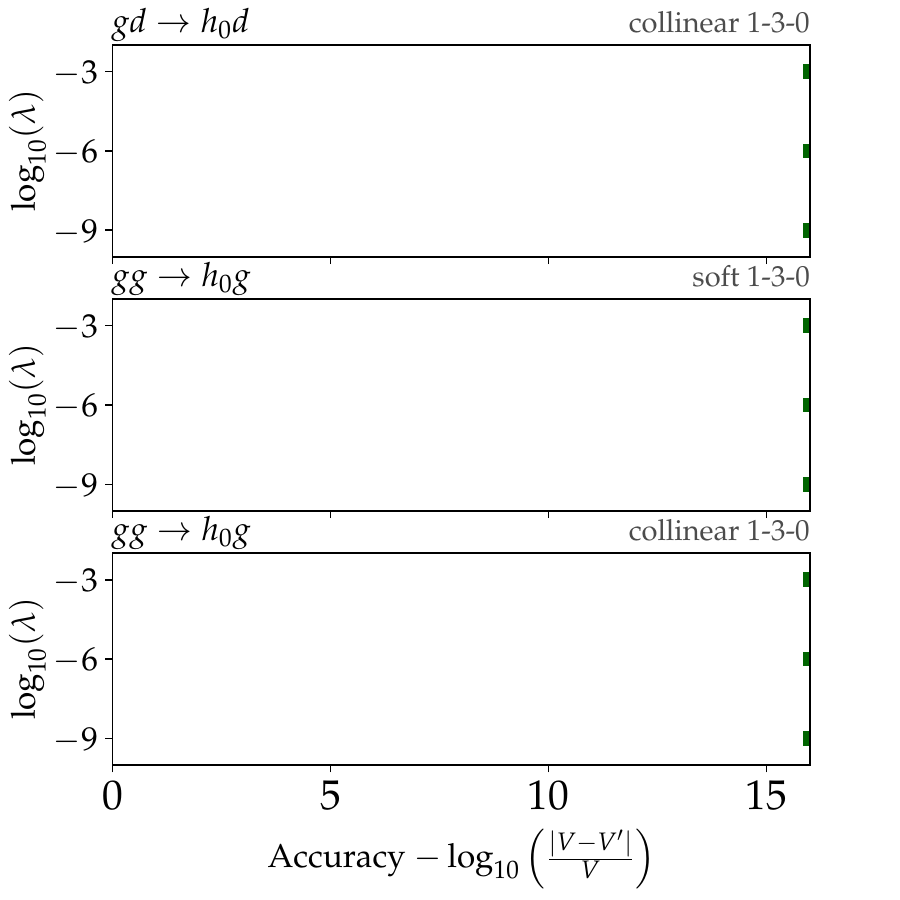}
  \includegraphics[width=0.32\textwidth]{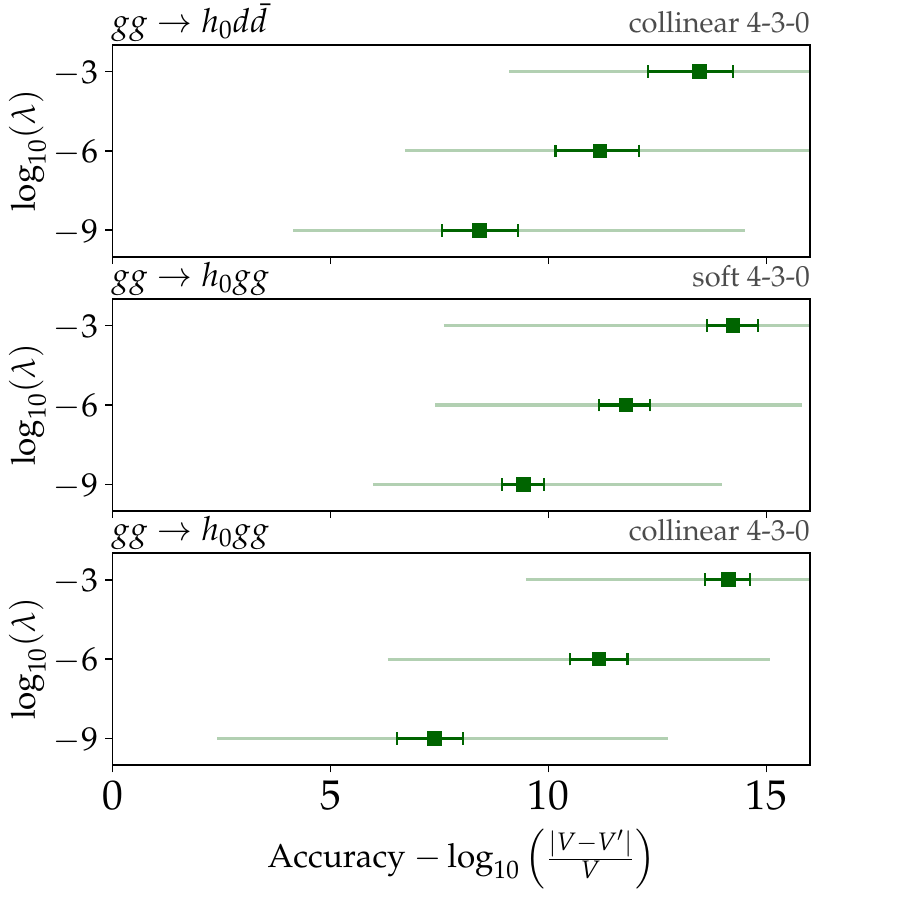}
  \caption[]{Test of the numerical accuracy of HEFT three- and four-parton one-loop amplitudes in the soft and collinear limits. Note that the accuracy is set to 16 digits in the $h_0+j$ case, where the two results agree perfectly.}
  \label{fig:scalingtestHEFT}
\end{figure*}

We consider crossings of the processes listed in Tables~\ref{tab:SM_Procs} and \ref{tab:HEFT_Procs} such that only final-state singularities are considered. We have validated that the numerical accuracy is generally worse when approaching final-state singularities, so that we deem this simplification sufficient.
For each singular limit, we generate $10^4$ hard phase-space points with $\sqrt{s}=1~\tera e\volt$ using \Sherpa and, depending on the singular limit of interest, rescale the momenta according to Eq.~\ref{eq:rescaleColl} or Eq.~\ref{eq:rescaleSoft} with $\lambda \in \{10^{-3}, 10^{-6}, 10^{-9} \}$. The results are collected in Fig.~\ref{fig:scalingtestSM3j} to Fig.~\ref{fig:scalingtestHEFT}, where each point corresponds to the average numerical accuracy according to Eq.~\ref{eq:numAcc} and the solid error bars indicate the $25\%$ quantiles of the median. The lighter-shaded error bands span from the worst to the best result in each run. In cases where the two results agree perfectly within machine precision\footnote{This is only the case for $pp\to h_0+j$.}, we set the number of stable digits to 16.

For all processes of interest, the numerical evaluation is sufficiently stable even in the deep infrared regions. Although not shown in the figures, we have checked that the stability of our interface is comparable to that of the other OLPs shown in Fig.~\ref{fig:runtimeRatioME} and Fig.~\ref{fig:runtimeRatioBVI} using appropriate settings.

\section{Conclusions}
We have presented a novel C++ interface to the well-known \MCFM parton-level Monte Carlo generator, giving access to its extensive library of analytical one-loop amplitudes.
The interface is generic and not tied to any specific Monte Carlo event generation tool. As a proof of its generality, we have implemented the interface in both, the \Sherpa and \Pythia event generators. The \Sherpa interface will become public with version 3.0.0, and the \Pythia interface is foreseen to become public in a future release of the 8.3 series\footnotemark[\vncfn]. It should be straightforward to adapt our code to the needs of other event generators.

We expect the interface to be valuable in two respects. First, for many of the processes considered here
the speedup over other OLPs is substantial; accessing these matrix elements via this interface rather than an
automated tool will therefore provide an immediate acceleration of event generation for many processes of
high phenomenological interest.  Second, the speed comparisons presented here highlight processes that are particularly
computationally intensive for automated tools. Further improvements to the efficiency of these codes may be possible,
with potential gains across a wider range of processes.  

The structure of the interface allows for simple extensions. Further one-loop matrix elements in \MCFM,
implemented either currently or in the future, may become accessible in a straightforward manner.
In the same spirit, the interface could also be extended to provide tree-level or two-loop matrix elements
included in \MCFM as the need arises.  Further extensions to the interface, for instance to provide finer control
over the one-loop matrix elements via the selection of helicities or color configurations, would also be possible.

Given that we have interfaced three popular automated OLPs within the generator-agnostic structure of the
new \MCFM interface, it is natural to envision the future development of a hybrid program that makes use of the
fastest matrix element library for each process.
Thinking further ahead, it may be worthwhile to reconsider a streamlined event generation framework, combining different (dedicated) parton-level and
particle-level tools. This idea has been pursued with \textsc{ThePEG} \cite{Lonnblad:2006pt}, but so far rarely deployed.
Apart from obvious efficiency improvements through the use of dedicated tools for different applications, such a framework enables previously unavailable methods for systematics studies.
In view of both the faster integration in \MCFM over \Sherpa and the magnitude of uncertainties pertaining to theoretical modeling of collider observables, this is becoming an increasingly important avenue for future work.

We want to close by highlighting that only a relatively small number of analytical amplitudes has to be known in order to cover a wide range of physical processes. When judiciously assembled, many parts of the calculations can be recycled in a process-independent way, with only charge and coupling factors being process-specific.
Compared to other efforts to increase the efficiency of event generators, swapping automated for analytical matrix elements is straightforward and simple. Analytical matrix element libraries provide a so-far little explored path towards higher-efficiency event generation for the (high-luminosity) LHC and future colliders.  

\section*{Acknowledgments}
We are grateful to Frank Krauss and Marek Sch{\"o}nherr for many fruitful discussions. We would also like to thank Stefano Pozzorini, Jonas Lindert, and Jean-Nicolas Lang for help with \OpenLoops, as well as Valentin Hirschi for help with \MadLoop.
CTP thanks Peter Skands for support. CTP is supported by the Monash Graduate Scholarship,
the Monash International Postgraduate Research Scholarship, and the J.~L.~William Scholarship.
This research was supported by the Fermi National Accelerator Laboratory (Fermilab), a U.S. Department of Energy, Office of Science, HEP User Facility.
Fermilab is managed by Fermi Research Alliance, LLC (FRA), acting under Contract No. \newline DE-AC02-07CH11359.
We further acknowledge support from the Monash eResearch Centre and eSolutions - Research Support Services through the MonARCH HPC Cluster.
This work was also supported in part by the European Union’s Horizon 2020 research and innovation programme
under the Marie Sk{\l}odowska-Curie grant agreement No 722104 – MCnetITN3.

\appendix

\onecolumngrid
\section{Parameters and cuts for timing comparisons}
\label{app:cuts_for_bvi}
In order to perform the timing comparisons shown in Fig.~\ref{fig:runtimeRatioBVI} and Table~\ref{tab:intTimesSherpaMCFM}.
we employ the following scale choices and phase-space cuts:
\begin{itemize}
\item $\Delta R_{\ell,\gamma}>0.4$
\item $\Delta R_{\gamma,\gamma}>0.4$
\item $p_{\mathrm{T},\gamma}>30~\mathrm{GeV}$
\item $p_{\mathrm{T},j}>30~\mathrm{GeV}$
\item $66~\mathrm{GeV}<m_{\ell\bar{\ell}}<116~\mathrm{GeV}$
\end{itemize}
We reconstruct jets using the anti-$k_\mathrm{T}$ algorithm~\cite{Cacciari:2008gp}
in the implementation of FastJet~\cite{Cacciari:2011ma}
with an $R$ parameter of 0.4. For the di-jet process we
require $p_{\mathrm{T},j}>$80~GeV.
Photons are isolated from QCD activity based on Ref.~\cite{Frixione:1998jh}
with $\delta_0$=0.4, $n$=2 and $\epsilon_\gamma$=2.5\%

\section{Structure of the interface}
\label{sec:app}
The \MCFM C++ interface is constructed as a C++ class
\begin{lstlisting}
CXX_Interface mcfm;
\end{lstlisting}
included in the header:
\begin{lstlisting}
#include "MCFM/CXX_Interface.h"
\end{lstlisting}
It must be initialized on a \texttt{std::map} of \texttt{std::string}s, containing all (standard-model) parameters:
\begin{lstlisting}
bool CXX_Interface::Initialize(
	std::map<std::string,std::string>& parameters);
\end{lstlisting}
Prior to use, each process has to be initialized in the interface:
\begin{lstlisting}
int CXX_Interface::InitializeProcess(const Process_Info &pi);
\end{lstlisting}
which takes a \texttt{Process\_Info} object as input, which in turn contains the defining parameters of a given process, i.e., the PDG IDs, number of incoming particles, and QCD and EW coupling orders:
\begin{lstlisting}
Process_Info(const std::vector<int> &ids, const int nin,
	const int oqcd, const int oew);
\end{lstlisting}
Phase space points are defined using the \texttt{FourVec} struct, which represents four-vectors in the ordering $(E, p_x, p_y, p_z)$.
\begin{lstlisting}
FourVec(double e, double px, double py, double pz);
\end{lstlisting}
Given a list of four-vectors in this format, one-loop matrix elements can be calculated either using the process ID returned by the \texttt{InitializeProcess} method
\begin{lstlisting}
void CXX_Interface::Calc(int procID,
	const std::vector<FourVec> &p, int oqcd);
\end{lstlisting}
or using a \texttt{Process\_Info} struct:
\begin{lstlisting}
void CXX_Interface::Calc(const Process_Info &pi,
	const std::vector<FourVec> &p,int oqcd);
\end{lstlisting}
In the same way, the result of this calculation can be accessed either via the process ID
\begin{lstlisting}
const std::vector<double> &CXX_Interface::GetResult(int procID);
\end{lstlisting}
or using the \texttt{Process\_Info} struct:
\begin{lstlisting}
const std::vector<double> &CXX_Interface::GetResult(const Process_Info &pi)
\end{lstlisting}
The result is returned as a list of Laurent series coefficients in the format $(\Order{\varepsilon^0}, \Order{\varepsilon^{-1}}, \Order{\varepsilon^{-2}}, \mathrm{Born})$. However, by default only the $\Order{\varepsilon^0}$ coefficient, i.e., the finite part, is returned. 
The calculation of the pole terms and the Born can be enabled by setting the following switch to 1:
\begin{lstlisting}
void CXX_Interface::SetPoleCheck(int check);
\end{lstlisting}

An example code showing the basic usage of the interface as well as a function filling the complete list of parameters with default values is publicly available\footnotemark[\tmpfn].
\twocolumngrid

\bibliographystyle{spphys}
\bibliography{bibliography.bib}

\end{document}